\documentclass[12pt]{iopart}

\usepackage{epsfig}

\begin{document}

\title{A global description of heavy-ion collisions}

\author{Thorsten Renk}
\address{Department of Physics, Duke University,
PO Box 90305,  Durham, NC 27708 , USA}
\ead{trenk@phy.duke.edu}


\begin{abstract}
A model for the evolution of ultrarelativistic heavy-ion collisions at both  
CERN SPS and RHIC top energies is presented. Based on the assumption of thermalization and 
a parametrization of the space-time expansion of the produced matter, this 
model is able to describe a large set of observables
including hadronic momentum spectra, correlations and abundancies, the emission of
real photons, dilepton radiation and the suppression pattern of charmonia. Each of these
obervables provides unique capabilities to study the reaction dynamics and taken together
they form a strong and consistent picture of the evolving system. Based on the emission
of hard photons measured at SPS, we argue that a strongly interacting, hot and dense system with temperatures
above 250 MeV has to be created early in the reaction. Such a system is bound to be
different from hadronic matter and likely to be a 
quark-gluon plasma, and we find that this assumption is in line with the subsequent
evolution of the system that is reflected in other observables.
\end{abstract}

\section{Introduction}

Lattice simulations (e.g.~\cite{lat1}) predict that  QCD
undergoes a phase transformation at a temperature $T_C \approx 150 - 170$ 
MeV \cite{lat2,lat3} from a confined hadronic phase to a
phase, the quark-gluon plasma (QGP), in which quarks and gluons constitute the relevant degrees of freedom
and the chiral condensate vanishes. 
Experimentally, this prediction can only be tested in ultrarelativistic
heavy-ion collisions. However, finding evidence, and ultimately proof, for the 
creation of a quark-gluon plasma faces several difficulties. Arguably the greatest challenge
is to link experimental observables to quantities measured on the lattice.

For a large set of observables, the evolution of the expanding medium is a key
ingredient for their theoretical description. It is also the place where results
from lattice QCD fit in: assuming a thermalized system is created,
its evolution is governed by the equation of state (EoS). 
Hence, in order to test the lattice QCD predictions, one has to start with this assumption
and show that it leads to
a good description of the experimental data. In the following, we will demonstrate that
this is indeed possible by introducing a parametrized evolution model.

\section{The model framework}

Assuming that an equilibrated system is created a proper time of order 0.5-1 fm/c after
the onset of the collision, we make the ansatz 
\begin{equation}
s(\tau, \eta_s, r) = N R(r,\tau) \cdot H(\eta_s, \tau)
\end{equation}
for the entropy density $s$ at given proper time
$\tau $ as measured in a frame co-moving with a given volume element, 
$\eta_s = \frac{1}{2}\ln (\frac{t+z}{t-z})$ the spacetime
rapidity and $R(r, \tau), H(\eta_s, \tau)$ two functions describing the shape of the distribution
and $N$ a normalization factor. We use Woods-Saxon distributions 
\begin{equation}
R(r, \tau) = 1/\left(1 + \exp\left[\frac{r - R_c(\tau)}{d_{ws}}\right]\right)
\end{equation}
and 
\begin{equation}
H(\eta_s, \tau) = 1/\left(1 + \exp\left[\frac{\eta_s - H_c(\tau)}{\eta_{ws}}\right]\right).
\end{equation}
for the shapes. Thus, the ingredients of the model are the skin thickness 
parameters $d_{ws}$ and $\eta_{ws}$
and the para\-me\-tri\-zations of the expansion of 
the spatial extensions $R_c(\tau), H_c(\tau)$ 
as a function of proper time. From the distribution of entropy density, the thermodynamics can be inferred via the EoS (as obtained in lattice calculations) and the particle emission function is then calculated using the 
Cooper-Frye formula for a freeze-out hypersurface
characterized by a temperature $T_F$. 
The total entropy is an input parameter whereas the entropy density
carried by the different hadronic degrees of freedom is calculated in a statistical
hadronization framework. The model is described in greater detail in \cite{SPS, RHIC}.

An important feature of the model is that it does not require boost invariance
but allows for an accelerated longitudinal expansion pattern instead with initial
rapidity $\eta_0$ not equal to final rapidity $\eta_f$. This implies
that in general the spacetime rapidity $\eta_s$ at some point is not equal 
to the rapidity $\eta$, rather there's a non-trivial relation dependent
on the trajectory leading to this point. In most relevant cases, however, the relation
can be approximated by $\eta = const. \cdot \eta_s$ \cite{SPS}. Such initial longitudinal compression
and re-expansion implies longer lifetime and higher initial temperature and energy density
as compared to a Bjorken scenario.

The adjustible parameters of the model are freeze-out temperature $T_F$, 
the radial expansion velocity $v_\perp$, the initial longitudinal expansion rapidity
$\eta_0$ and the skin thickness parameters $d_{ws}, \eta_{ws}$.

\section{Hadronic observables}

\begin{figure}[htb]
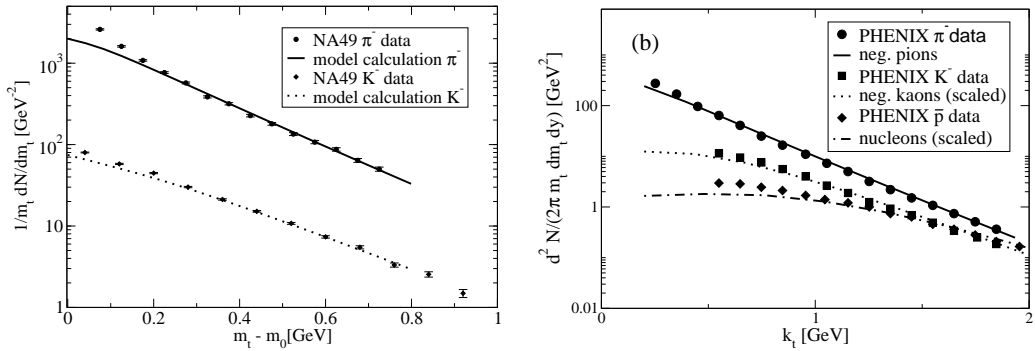

\begin{center}
\vspace*{0.5cm}
\epsfig{file=m_t-sps.eps, width=6.6cm} \hspace{0.2cm}
\epsfig{file=spectra_rhic.eps, width=6.6cm} 
\end{center}
\caption{\label{F-SPEC}Transverse mass spectra obtained by NA49 \cite{NA49-1}
(left) and by PHENIX \cite{Phenix-Spectra} (right) compared with model results. The model does not
include contribution from resonance decays relevant at low transverse mass.}
\end{figure}

\begin{figure}[htb]
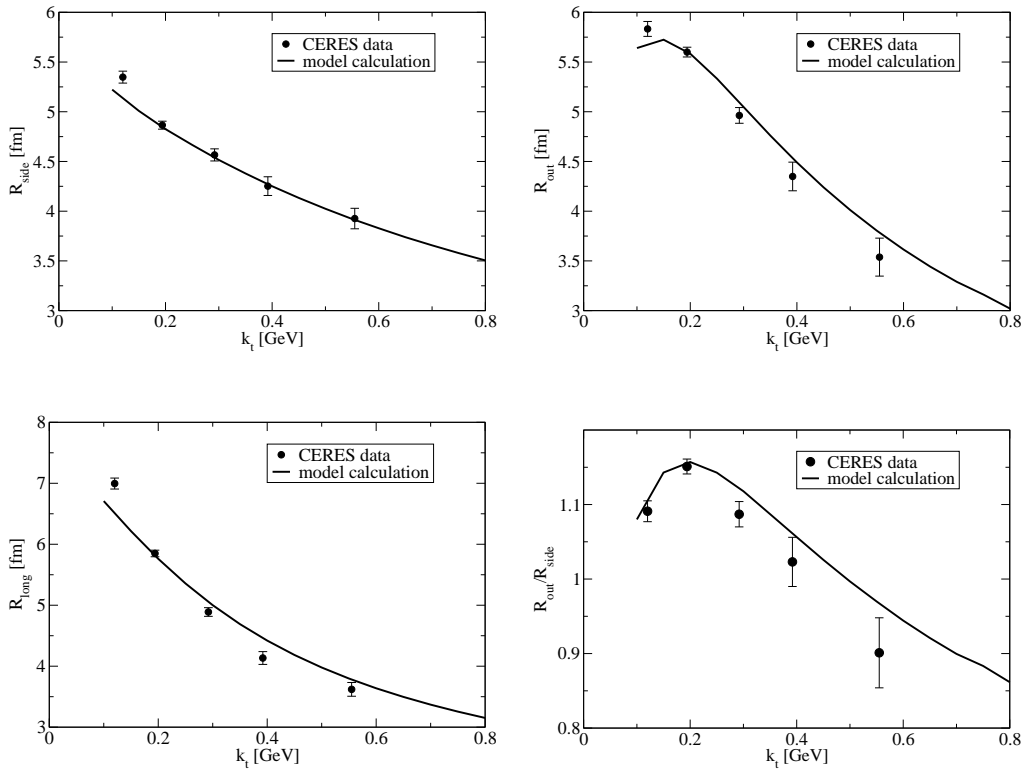

\begin{center}
\vspace*{0.5cm}
\epsfig{file=R_side.eps, width=6.5cm} \hspace{0.2cm}
\epsfig{file=R_out.eps, width=6.5cm} 

\vspace*{0.8cm}
\epsfig{file=R_long.eps, width=6.5cm} \hspace{0.2cm}
\epsfig{file=R_ratio.eps, width=6.5cm}
\end{center}
\caption{\label{F-HBT} HBT Correlation radii obtained by the CERES collaboration
\cite{CERES-HBT} as a function of transverse pair momentum $k_t$ 
compared with the model results. No systematic errors are included
in the experimental errorbars.}
\end{figure}

\begin{figure}[htb]
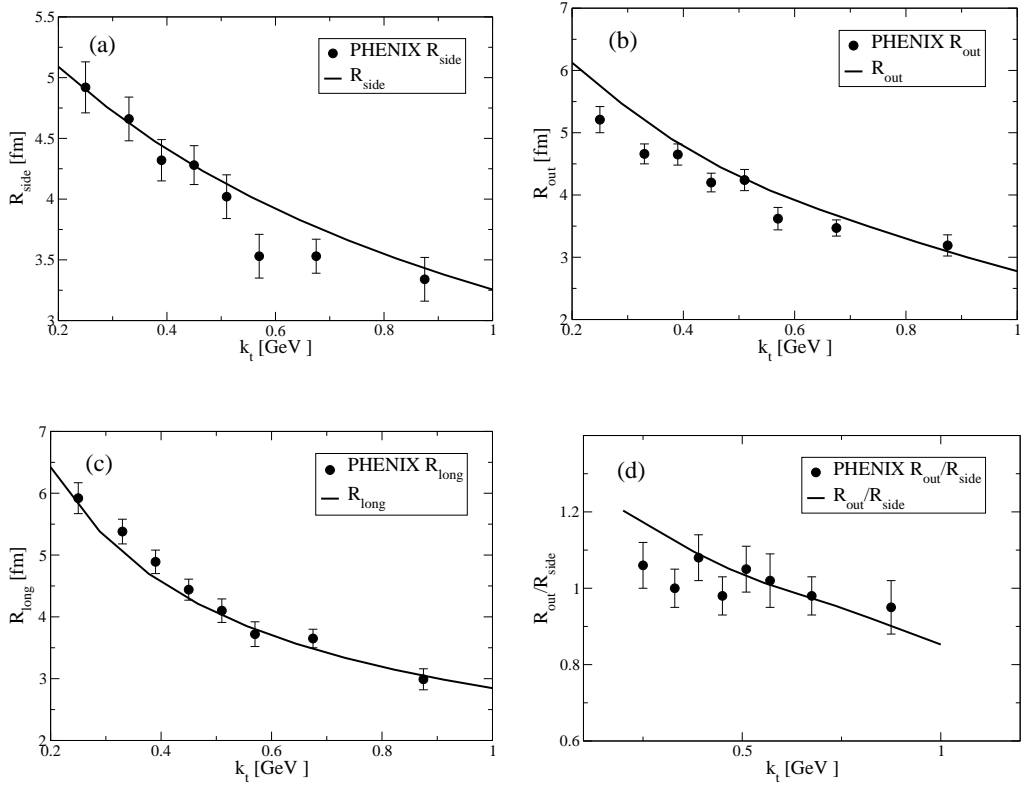

\begin{center}
\vspace*{0.5cm}
\epsfig{file=R_side_rhic.eps, width=6.5cm} \hspace{0.2cm}
\epsfig{file=R_out_rhic.eps, width=6.5cm} 

\vspace*{0.8cm}
\epsfig{file=R_long_rhic.eps, width=6.5cm} \hspace{0.2cm}
\epsfig{file=R_ratio_rhic.eps, width=6.5cm}
\end{center}
\caption{\label{F-HBT_RHIC} HBT Correlation radii obtained by the PHENIX collaboration
\cite{Phenix-HBT} as a function of transverse pair momentum $k_t$ 
compared with the model results.}
\end{figure}

In the following, we only focus on central collisions.
The model parameters are fitted to the recent SPS data for
158 AGeV Pb-Pb collisions obtained by NA49 and CERES \cite{NA49-1, CERES-HBT}
and 200 AGeV Au-Au collision data obtained by the PHENIX collaboration at RHIC
\cite{Phenix-Spectra, Phenix-HBT}. The data for both SPS and RHIC favour
a comparatively low freeze-out temperature of order 100-110 MeV and
a strong transverse flow ($v_\perp = 0.57$ at SPS and $0.65$ at RHIC).
In both cases, the initial rapidity inverval is found to be significantly
smaller than the finally observed rapidity interval $\eta_f$, we find
$\eta_0 = 0.55$, $\eta_f = 1.45$ at SPS and $\eta_0 = 1.8, \eta_f = 3.6$ at
RHIC. This is a crucial ingredient for the successful description of the data and the 
numbers indicate that the actual dynamics of the system could
be very different from a Bjorken expansion.
The resulting transverse mass ($m_t$) spectra for different hadron species and the pionic
Hanbury-Brown Twiss (HBT) correlation radii show good agreement with the data (Figs.~\ref{F-SPEC}, \ref{F-HBT} and
\ref{F-HBT_RHIC}).

There is a contribution to the $\pi^-$ $m_t$-spectra 
coming from the vacuum decay of resonances after kineitc decoupling at
$\tau_f$. The model is able to calculate the magnitude
of this contribution using statistical hadronization but not
its $m_t$ distribution, therefore it is left out when we compare with data.
This explains the mismatch at low $m_t$ and that the
integrated spectrum does not yield the multiplicity indicated by the data.
The same is visible in the K$^-$ spectra and the nucleon spectra, 
albeit less pronounced.

\section{Electromagnetic observables}

\begin{figure}[htb]
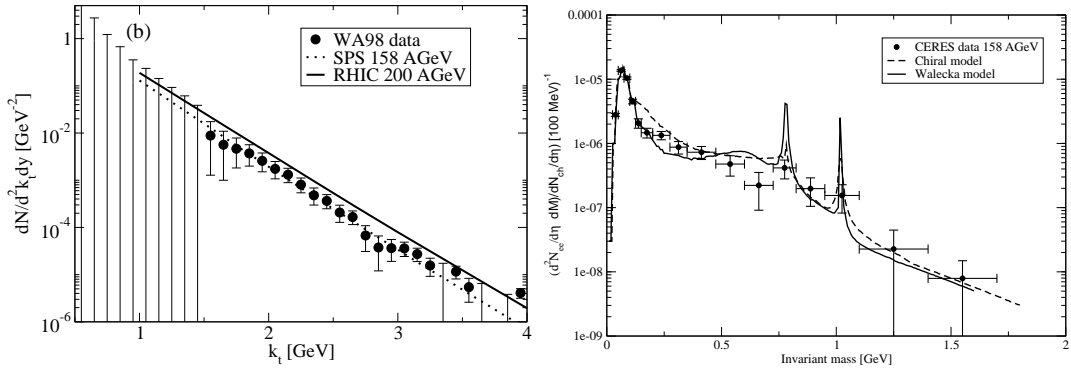

\begin{center}
\vspace*{0.5cm}
\epsfig{file=photon_rhic_new.eps, width=7cm}
\epsfig{file=Chiral_vs_Hadronic.eps, width=7.0cm}
\end{center}
\caption{\label{F-EM}Left: The direct photon spectrum measured at the CERN
SPS \cite{PhotonData} as compared with the model calculations for SPS (dotted) and
RHIC (solid). Right: The dilepton spectrum compared with SPS 158 Pb-Au
results \cite{CERES} (dots) using both the chiral model (dashed) and the 
mean field model (solid)
spectral function. }
\end{figure}

For the emission rate of direct photons, we use the complete $O(\alpha_s)$ calculation
\cite{2-2-Kapusta,2-2-Baier,Aurenche1, Aurenche2, Aurenche3,Complete1, Complete2} in the
form of the parameterization provided in \cite{Complete2}.
The spectrum of emitted photons can be found by folding the rate \cite{Complete2} with the
fireball evolution. In order to account for flow, the energy of a photon emitted
with momentum $k^\mu =(k_t, {\bf k_t}, 0)$ has to be evaluated in the local rest
frame of matter, giving rise to a product $k^\mu u_\mu$ with $u_\mu(\eta_s, r, \tau)$
the local flow profile.
For comparison with the measured photon spectrum \cite{PhotonData}, we present the 
differential emission spectrum into the midrapidity slice $y=0$. 
The resulting photon spectrum is shown in Fig.~\ref{F-EM} (left). Details of the calculation
can be found in \cite{SPS, Photons, Photons_RHIC}.

The lepton pair emission rate from a hot domain populated by particles in
thermal equilibrium at temperature $T$ is proportional to the imaginary part
of the spin-averaged photon self-energy, with these particles as
intermediate states. The thermally excited particles annihilate to yield a
time-like virtual photon with four-momentum $q$ which decays subsequently into
a lepton-antilepton pair. The information about the strong interaction dynamics is 
encoded in the spectral function which enters the emission rate. 
For its computation, we have to use different techniques for dealing with
partonic and hadronic degrees of freedom. In the partonic phase, we use a leading
order calculation using thermal quasiparticles as degrees of freedom. 
In the hadronic phase we calculate in two different approaches to estimate the
theoretical uncertainty - a chiral model 
\cite{Dileptons} and a mean-field model \cite{Amruta}. 

The resulting invariant mass spectrum agrees with the measured data for both the
chiral model and the mean field model of hadronic matter.
For SPS, we find photons in the momentum range between 2-3 GeV to be dominated by
thermal emission from an initial high-temperature partonic phase whereas the dilepton
yield in the invariant mass region below 1 GeV is dominated by emission from hadronic
matter. Both quantities are sensitive to the 4-volume of radiating matter and
complement each other in supporting the fireball evolution model in its different evolution phases.
Most notably, the dilepton result indicates that hadronic matter does not cease to
interact after the phase transition, instead substantial modifications of the
$\rho$ properties in medium are required by the data. The photon spectrum cannot be
reproduced in a Bjorken expansion, thus confirming the findings outlined above based
on an analysis of hadronic observables. The sensitivity of the yield to the
equilibration time $\tau_0$ allows to obtain $\tau_0 < 3$ fm/c \cite{SPS, Photons}.

For RHIC, we find that the contribution to the photon spectrum coming from hadronic
matter is as large as the yield from partonic matter due to the strong flow. This
is unfortunate, as it reduces the capability of a photon measurement to help
in the determination of the peak temperature reached in the collision \cite{Photons_RHIC}.
In both calculations, a pre-equilibrium contribution to the photon yield is
missing. Such a contribution is expected to be most relevant for the high momentum
region and to dominate the yield above 3-4 GeV. 

\section{Charmonium}

\begin{figure}[htb]
\begin{center}
\epsfig{file=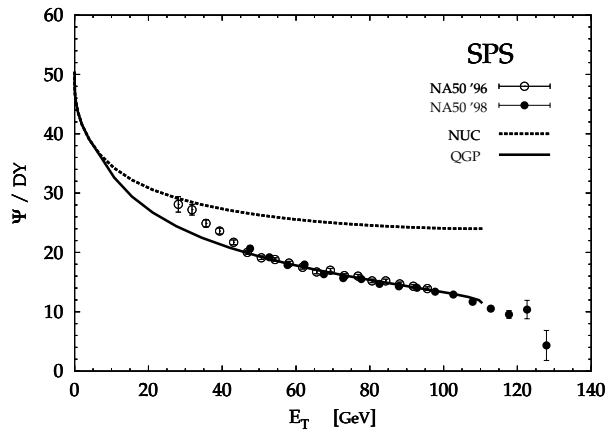,width=5.8cm,angle=-90}
\end{center}
\caption{\label{F-CHARM}Result at SPS energy for the ratio $\Psi/DY$ as function 
of the transverse energy. The dashed curve includes only
nuclear effects, while the full line is the complete result including gluon 
dissociation.}

We find good agreement with the data with the exception of the high $E_T$ region where
fluctuations in the transverse energy (not included in the model) are relavent and
in the low $E_T$ region where the system cannot be expected to be thermalized any more.

\end{figure}

Charmonia yields are calculated by solving rate equations for the interaction of the state with the
medium. The initial condition  for the rate equations 
is determined by comparing to the production in p-p and p-A collisions.
For SPS conditions, dissociation of the state dominates and we find the equation 

\begin{equation}
\frac{d}{d\tau} N^y_\Psi(\tau) = - \lambda_D(\tau)\,N^y_\Psi(\tau)
\,, \quad \lambda_D(\tau) = \sum_n\ \langle\langle\,\sigma^n_D\,v_{rel}
\,\rangle\rangle(\tau)\ \rho_n(\tau) 
\end{equation}
for the rapidity density  $N^y_\Psi(\tau)$ as a function of proper time $\tau$
where $\rho_n$ is the medium density \cite{SPS,Charm}. Folding this equation with
the fireball evolution yields Fig.~\ref{F-CHARM}.

\section{Summary}

We have presented a thermal description of ultrarelativistic heavy-ion collisions at the CERN SPS
for 158 AGeV Pb-Pb which is highly consistent and leads to a good
description of a large set of different observable quantities. 
While this does not provide a conclusive  proof for the creation of a thermalized system (and hence a
QGP), it constitutes certainly strong evidence for it. It remains to be investigated if the results
can be reproduced in a thermal microscopical transport description (like hydrodynamics) and
if a non-thermal framework is able to describe the experimental results as well or if thermalization is
required by the data.

For RHIC, we have presented a fireball evolution model which is able to simultaneously
describe both HBT correlations and transverse mass spectra. This parametrized
evolution model will hopefully be able to guide hydrodynamical models (which currently
fail to provide such a simultaneous description).  
We are now in the process of making predictions
for other observables based on this evolution model.

\ack

I would like to thank B.~M\"{u}ller, S.~A.~Bass
for interesting discussions, helpful comments and support.
This work was supported by the DOE grant DE-FG02-96ER40945 and a Feodor
Lynen Fellowship of the Alexander von Humboldt Foundation.

\end{document}